\documentclass[12pt, a4paper]{article}

\usepackage[left=1in,right=1in,top=1.0in,bottom=1in]{geometry}
\usepackage{amsmath,amssymb,amsthm,amsfonts,afterpage,hyperref}
\usepackage{amscd,subeqnarray}
\usepackage{graphicx}
\usepackage{caption}
\usepackage{subcaption}
\usepackage{authblk}
\usepackage{tikz}
\usepackage{array}
\usepackage{wrapfig}
\usepackage{color}
\usepackage{ragged2e}
\usepackage{stmaryrd}
\usepackage{siunitx}
\usepackage{multirow}
\usepackage{xcolor,cancel}
\usepackage{boldfonts}
\usepackage{symbols}
\usepackage{footmisc}
\usepackage{float}
\usepackage{setspace}
\usepackage{bm}
\usepackage{booktabs}
\usepackage{tikz, xcolor}
\usepackage{soul}
\usepackage[T1]{fontenc}
\usepackage[utf8]{inputenc}
\usepackage[sort, numbers]{natbib}
\setcitestyle{square}
\usetikzlibrary{shapes,arrows, positioning,calc}	
\usetikzlibrary{arrows,snakes,backgrounds}

\usepackage{algorithm}
\usepackage{algpseudocode}

\DeclareMathAlphabet{\mathpzc}{OT1}{pzc}{m}{it}

\newtheorem{definition}{Definition}
\newtheorem{theorem}{Theorem}[section]

\newtheorem{lemma}[theorem]{Lemma}
\newtheorem{remark}{Remark}

\providecommand{\keywords}[1]
{
  \small	
  \textbf{\textit{Keywords---}} #1
}

\title{Analyzing mob dynamics in social media networks using epidemiology model}

\author[1]{Ahmed AL-Taweel}
\author[2]{Saqib Hussain}
\author[3]{S. M. Mallikarjunaiah}

\affil[1]{Mathematical Sciences, Georgia Southern University, Statesboro, GA 30458, USA}
\affil[2]{Department of Mathematics and Physics, Texas A\&M International University, Laredo, TX 78041}
\affil[3]{Department of Mathematics \& Statistics, Texas A\&M University-Corpus Christi, Corpus Christi, TX 78412, USA}

\date{} % Suppresses the date

\begin{document}

\maketitle

\thispagestyle{empty} % Suppresses page number on the title page

\vspace{-2em} % Adjust vertical space as needed
\begin{center}
\small
\textsuperscript{1}aaltaweel@georgiasouthern.edu \\
\textsuperscript{2}saib.hussain@tamiu.edu \\
\textsuperscript{3}m.muddamallappa@tamucc.edu
\end{center}
%---------------------------------------------------------------------
\begin{abstract}
Epidemiological models, traditionally used to study disease spread, can effectively analyze mob behavior on social media by treating ideas, sentiments, or behaviors as ``contagions" that propagate through user networks.  In this research, we introduced a mathematical model to analyze social behavior related to COVID-19 spread by examining Twitter activity from April 2020 to June 2020. Our analysis focused on key terms such as ``lockdown" and ``quarantine" to track public sentiment and engagement trends during the pandemic. The threshold number $\Re_{0}$ is derived, and the stability of the steady states is established. Numerical simulations and sensitivity analysis of applicable parameters are carried out. The results show that negative sentiment on Twitter has less influence on COVID-19 spread compared to positive sentiment. However, the effect of negative sentiment on the spread of COVID-19 remains remarkably strong. Moreover, we use the Caputo operator with different parameter values to study the impact of social media platforms on the transmission of COVID-19 diseases.  
\end{abstract}
\noindent \keywords{Mob, flash mob, COVID-19 pandemic, Social behavior,  Equilibrium, Stability, epidemiological modeling, Twitter, Online social networks}

%---------------------------------------------------------------------------------
\section{Introduction}\label{Section:Introduction} 
The event of mob propagation has been studied across numerous fields, including sociology, communication studies, and computational modeling. In \cite{gustave2018crowd}, it is explained how individuals in a crowd can behave differently than they would on their own, with anonymity and emotional contagion playing crucial roles. Recently, in \cite{le2023crowd}, authors developed theories for the digital world, showing how these factors still impact behavior online.
%\textcolor{red}{The phenomenon of mob propagation has been examined across multiple disciplines, including sociology, psychology, communication studies, and computational modeling. In (Gustave Le Bon's "The Crowd: A Study of the Popular Mind,), Le Bon laid the foundation for understanding how individuals in a crowd can exhibit behavior that deviates from their normal actions. Authors highlighted the role of anonymity and emotional contagion in facilitating crowd behavior, which modern researchers have extended to digital spaces.}

Social media has increasingly become a double-edged sword in facilitating constructive and destructive group mobilization forms. On the one hand, platforms can unite people for creative or community-oriented activities, but on the other hand, they can trigger harmful mob events. The \textbf{UK Anti-Migrant Riots} in August 2024 illustrates how misinformation spread through social media can lead to widespread chaos. False claims about a stabbing incident fueled far-right protests across the UK, escalating into violent anti-migrant demonstrations and multiple arrests. These incidents highlight the potent role of social media in amplifying mob behavior, emphasizing the critical need for robust content moderation and the responsible use of these platforms to mitigate their misuse.

During the COVID-19 pandemic, Twitter played a pivotal role in shaping public discourse and mobilizing collective behavior, both supporting and opposing public health measures. While many users leveraged the platform to share accurate health information and promote adherence to safety protocols such as social distancing and \textbf{lockdowns}, a significant wave of resistance emerged, fueled by misinformation, skepticism, and political motivations. Hashtags like \textbf{\#EndTheLockdown}, \textbf{\#ReopenAmerica}, and \textbf{\#FreedomOverFear} trended widely, organizing anti-lockdown protests and amplifying conspiracy theories. This mob behavior frequently translated into real-world actions, including protests that defied social distancing mandates and escalated into confrontational encounters. Furthermore, the spread of misinformation undermined public health messaging, complicating efforts to maintain compliance with safety measures. These dynamics highlighted Twitter’s dual role as both a disseminator of valuable information and a tool for mobilizing resistance, underscoring the critical need for digital literacy, content moderation, and robust public health communication strategies to mitigate the adverse impacts of social media during crises.

Understanding how information spreads on social media and its subsequent impact on public behavior is paramount. Epidemiological models, traditionally developed to study the spread of diseases, have been adapted to analyze social phenomena by treating ideas and sentiments as "contagions" propagating through user networks. These models allow researchers to capture the dynamics of user engagement, predict trends, and evaluate interventions. In this research, we apply an SEIAR (Susceptible-Exposed-Infected-Against-Recovered) model to analyze social behavior related to COVID-19, focusing on user activity on Twitter between April 2020 and June 2020. By leveraging this framework, we aim to quantify the spread of public sentiment, identify influential parameters, and evaluate their sensitivity to better understand the dynamics of online social movements.

 Epidemiological models, such as compartmental models, play a crucial role in organizing populations into distinct groups, providing a foundational mathematical framework for comprehending epidemic dynamics. \cite{24} introduced a fundamental SI model that classifies the population into susceptible and infected, extendable to an SIS model where individuals in the infected compartment can revert to a susceptible state \cite{25}. The SIR model, a widely recognized epidemic model \cite{26,25,luo2013finding,shi2008sis}, categorizes individuals into three compartments: susceptible, infected, and recovered. The SEIZ model (susceptible, exposed, infected, skeptic) has been employed to simulate the spread of news and rumors on the Twitter platform \cite{32,mathur2020dynamic,jin2013epidemiological}. Numerous scholars have used the infectious disease dynamics model for smoking, alcoholism, drug addiction, game addiction, social media addiction, rumor diffusion model in digital social networks and other issues \cite{alemneh2021mathematical,huo2016modeling,yang2024rumor,guo2023dynamic,tian2019rumor,zhu2020stability,brauer2019mathematical,gurung2024decoding,wewer,sun2024dynamic}. In this work, we applied a SEIAR (Susceptible-Exposed-Infected-Against-Recovered)  model to analyze social behavior related to COVID-19 by examining Twitter activity between April 2020 and June 2020. Our analysis focused on key terms such as ``lockdown" and ``social distancing" to track public sentiment and engagement trends during the pandemic.

 The paper is organized as follows. Section \ref{secM} presents our methodology, including Twitter data and a mathematical framework for the epidemiological model, and focuses on the various components contributing to our understanding of the spread of mob events over time. Section \ref{33} investigates model analysis, including the threshold number $\Re_{0}$, the model's stability analysis, and sensitivity analysis. The results from numerical analysis are reported in Section \ref{sec3} to validate the theoretical aspect of the model. Finally, concluding remarks and future research plans are given in Section \ref{sec4}.
%%%%%%%%%%%%%%%%%%%%
\section{Methodology}\label{secM}
This section outlines data collection and the methodology employed in this article.
%%%%%%%%%%%%%%%%%%%%
\subsection{Data Collection}
Social media networks (Twitter, Telegram, Facebook, Instagram, Facebook, etc.) have developed into sites of intensive and incessant information dissemination between government organizations and the general public due to the COVID-19 epidemic's global spread.   Numerous scientific research studies have indicated that social media platforms and other news websites may be helpful data sources for problem analysis and understanding individuals’ behavior over time during a pandemic \cite{world2022director,csse2020covid, gurung2023analyzing}. Numerous monitoring techniques have been performed to aid
public health management and specialists in making decisions to determine enormous volumes of data from social media platforms. This data can be utilized to detect the ideas, attitudes, emotions, sentiments, and topics on which individuals’ senses are concentrated in reaction to the COVID-19 virus \cite{liu2021public}. Systematic analysis of the data assists policymakers and health specialists in determining public concerns and resolving them most effectively. We collected COVID-19 discourse tweets on Twitter from April 2020 to June 2020. To obtain the tweets, we utilized a developer account on Twitter, which allowed us to access Twitter API v2 to explore tweets on specific subjects.  We could search Twitter's archive database of tweets based on keywords and date specifications, using the code representatives from Twitter Developer Relations and the search-tweets-python package to assist us in tweet collections (https://github.com/twitterdev/search-tweets-python). All tweets were collected utilizing a set of COVID-19-related keywords, including ${``COVID-19"}$,  ${``coronavirus"}$, ${``corona"}$, ${``quarantine"}$, ${``homequarantine"}$, ${``quarantinecenter"}$, ${``socialdistancing"}$, ${``staysafe"}$, ${``covid"}$, ${``covaccine"}$, ${``lockdown"}$, ${``stayhome"}$.  These keywords were selected based on their popularity on Google Trends about COVID-19 during data collection. The parameters for the tweet search contain data fields, for instance ${``tweet_{-}ID"}$,  ${``user_{-} mention"}$, ${``like_{-}count"}$, ${``quote_{-}count"}$, ${``creation_{-}data"}$, ${``favorite_{-}count"}$, ${``reply_{-}count"}$, ${``retweet_{-}count"}$, ${``source"}$, ${``tweet"}$, ${``hashtags"}$, ${``stayhome"}$. We preprocessed the collected data by setting a user-defined preprocessing operation based on NLTK (Natural Language Toolkit, a Python library for NLP). In the initial step, all the tweets are converted into lowercase. Then, it removes all additional white spaces, special characters, numbers, ASCII characters, URLs, punctuations, and stopwords from the tweets. Then, it restores all `COVID' words into `COVID-19', as we already cleared all numbers from the tweets. The pre-processing function reduces the number of inflected words to their word stem, utilizing stemming.
%%%%%%%%%%%%%%%%%%%%%
\subsection{The SEIAR Mob propagation model}
\begin{figure}[H]
	\centering
	\includegraphics[width=0.7\textwidth]{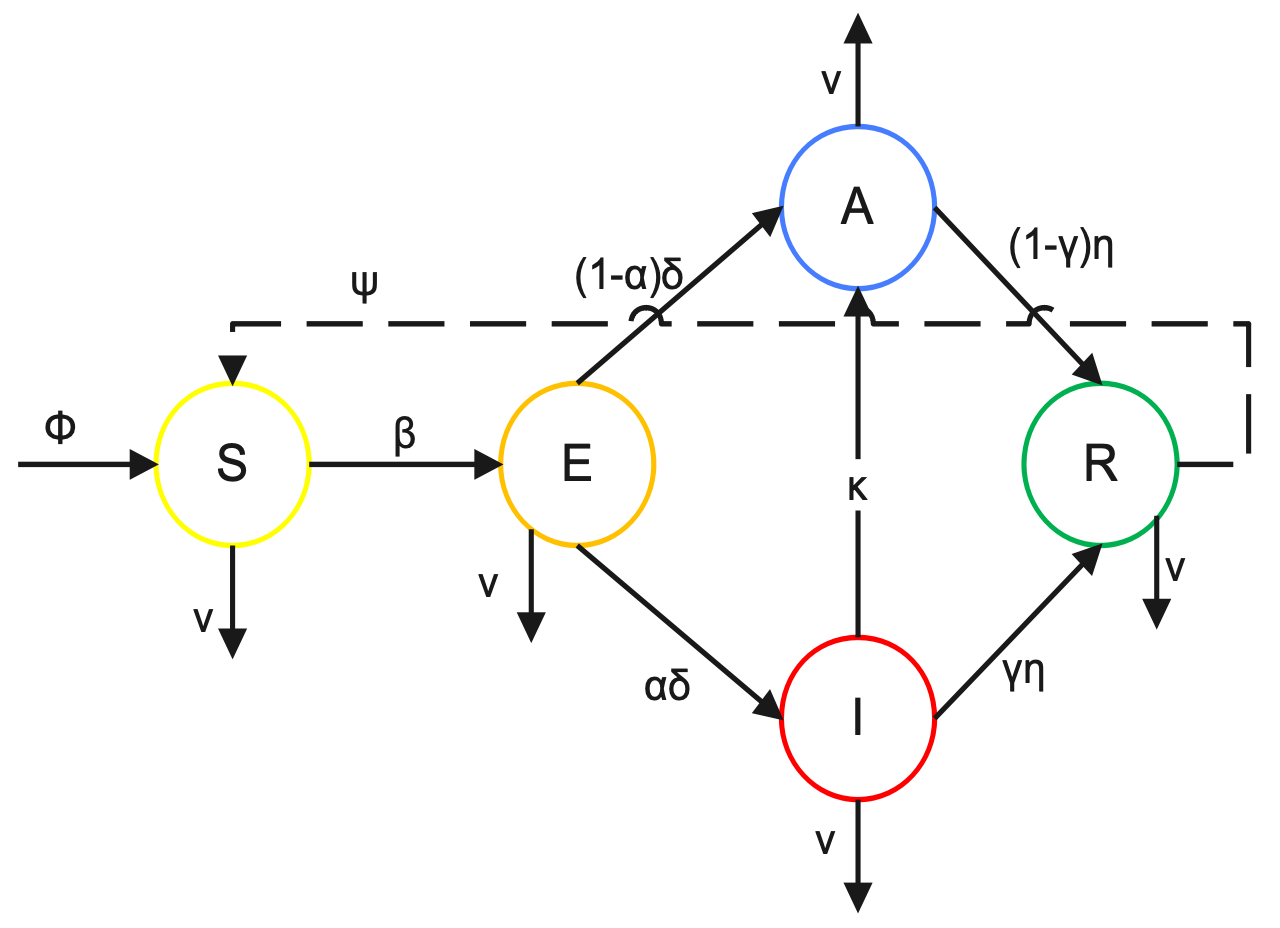} % Adjust the height as needed
	\caption{ Flow chart of mob propagation model \ref{eq:system22}.}%SEAGR  compartment model.}
	\label{SEIREq}
\end{figure}

A mob is an event managed through social media networks, email, or other digital communication technologies. A group of individuals contacts each other online or offline to conduct an act collectively, and then it disperses over a long period or fast \cite{al2024evaluating}. Studying such a mob phenomenon is challenging due to lacking data, theoretical techniques, and resources. For this purpose, we employ epidemiological techniques to model the behavior of mobbers. The total population is divided into five compartments: $S(t)$, the number of susceptible individuals who are not aware of the event but susceptible to Mob event by online social media platforms or offline; $E(t)$, the number of individuals who learned about the mob event but have yet to grow to the mob stage, waiting for a short delay before deciding whether to join in or not; $I(t)$, the Individual who is engaged in the event and spend most of their time on it; $A(t)$, the Individual who decide to go against the mobbers; $R(t)$, the Individual who disengaged or retrieved from the event and stopped sharing it.
 The five ordinary differential equations (ODEs) system  \ref{eq:system22} controls our model. A transfer diagram of the model is displayed in Fig. \ref{SEIREq}, and the model's parameters are shown in Table \ref{tab:seiz-parameters}.
 %%%%%%%%%%%%%%%%%%%%
{\small \begin{table}[h]
			\centering
			\caption{Definitions of Parameter  OF the SEIAR model \ref{eq:system22}.}
               \resizebox{1.1\textwidth}{!}{ % Adjust 0.8 to the desired width
			\label{tab:seiz-parameters}
				\begin{tabular}{ccc}
					\toprule
					Parameter & Definition&Value\\
					\midrule
					$\Phi$ & The joining number of susceptible individuals per unit of time&0.5\\
     					$\beta$ & The transmission rate of exposure to the susceptible individual&0.25\\
                 $\alpha$ & The ratio of exposed individuals that enter or disjoint to mobber class&0.05\\
      $\delta$ &  Individuals that exit exposed class&0.1\\
      k& Individuals who enter A class due to the treatment&0.05\\
      $\gamma$& Proportion of recovered individuals to mob event&0.1\\
      $\eta$&  Individuals who join the recovered class&0.1\\
                    $\nu$ &  The  emigration rate of moving out of the system after the population moves &0.05\\
                  $\Psi$ &  Individuals who leave the recovered class&0.02\\ 
                  
					\bottomrule
			\end{tabular}}
\end{table}}
%%%%%%%%%%%%%%%%%%%%%%%%%%%%
 \iffalse
\section{Model construction}
The SEAGR model comprises five compartments:
	\begin{itemize}
		\item[1.] Susceptible individuals (denoted by S) are those who are not aware of the event but susceptible to Mob event by onlne social media platforoms or offline. \\
		
		\item [2.] Exposed individuals (Denoted by E) are those who learned about the mob event but have yet to grow to the mob stage, waiting for a short delay before deciding whether to join in or not. \\
		
		\item[3.]  Mobbers individuals (denoted by A) are individuals who are engaged in the event and spend most of their time on it. \\
		
		\item[4.] Against the mobbers (denoted by G) are individuals who decide to go against the mobbers.\\
		
		\item [5.] Recovered individuals (denoted by R) are those individuals who disengaged or retrieved from the event and stopped sharing it. 
		
\end{itemize} 
\fi 

The following system of ODEs represents the SEIAR model :
\begin{equation}
			\begin{cases}
            	\begin{split}
				\frac{dS}{dt} &=\Phi+\Psi R-\beta S  E -\nu S,\\
				\frac{dE}{dt} &= \beta S E-\alpha \delta E-(1-\alpha)\delta E-\nu E, \\
				\frac{dI}{dt} &=\alpha \delta E-k I-\gamma \eta I-\nu I,\\
				\frac{dA}{dt} &= (1-\alpha)\delta E+k I-(1-\gamma)\eta A-\nu A,\\
				\frac{dR}{dt} &=(1-\gamma)\eta A+\gamma \eta I-\Psi R-\nu R.  
				\end{split}
            \end{cases}
			\label{eq:system22}
\end{equation}

The total number of individuals on social networks $N(t)$ satisfies the following equation:
$$N(t) =S(t) +E(t) +A(t) + G(t) + R(t),$$
with initial conditions $ S(0) \geq 0,  E(0) \geq 0 ,  I(0) \geq 0 ,  A(0) \geq 0 , \text{ and }  R(0) \geq 0 $ at $ t = 0 .$
Then, we have
	\begin{equation}
		\frac{dN}{dt}\leq \frac{\Phi}{\nu}.\label{eq11}
	\end{equation}
	The solution of the above equation is defined by $N(t)\leq N(0)\exp(-\nu t)+\frac{\Phi}{\nu}(1-\exp(-\nu t))$. Then, as $t\longrightarrow\infty, N(t)\longrightarrow\frac{\Phi}{\nu}$. Therefore, the model positively invariant region is presented by:
		\begin{equation}
	\varSigma=\left\{\left(S(t),E(t),I(t),A(t),R(t))\in R^{5}_{+}: 0< N(t)\leq \frac{\Phi}{\nu}\right)\right\}.
\end{equation}
%%%%%%%%%%%%%%%%%%%%%%%%%%%
\section{Model analysis}\label{33}
Now, we can present the following equilibrium points and basic reproduction numbers for the equations \ref{eq:system22}.
\subsection{Equilibrium points}
To get the equilibria, we seek the time-independent solutions $(S^{\prime}, E^{\prime}, I^{\prime}, A^{\prime}, R^{\prime})$  that satisfy system \ref{eq:system22} with the time derivatives set to zero. The resulting system is as follows:
\begin{equation}
			\begin{cases}
            	\begin{split}
				0 &=\Phi+\Psi R-\beta S  E -\nu S,\\
				0 &= \beta S E-\alpha \delta E-(1-\alpha)\delta E-\nu E, \\
				0 &=\alpha \delta E-k I-\gamma \eta I-\nu I,\\
				0 &= (1-\alpha)\delta E+k I-(1-\gamma)\eta A-\nu A,\\
				0 &=(1-\gamma)\eta A+\gamma \eta I-\Psi R-\nu R. 
                \end{split}
			\end{cases}
			\label{eeeq1}
\end{equation}
By solving the system of equations \ref{eeeq1}, in the absence of the social media mob, we assume that $E=I =A=R= 0$. Therefore, the mob-free equilibria (MFE) are given by:
\begin{equation}\label{MFE}
	MFE=\left(S_{0},0,0,0,0\right) \text{ where }  S_{0}=\frac{\Phi}{\nu}.
\end{equation}
%%%%%%%%%%%%%%%%%%%%%%%%%%%%%%%%%%%%%%%%%%%%%%%%
% Theoertical part 
%%%%%%%%%%%%%%%%%%%%%%%%%%%%%%%%%%%%%%%%%%%%%%%
Next, the threshold is computed by utilizing the next-generation matrix method.
\subsection{Basic reproduction number $\Re_{0}$}
In this section, we will use the basic reproduction number $\Re_0$ of the model \ref{eq:system22} to analyze the nature of mob propagation. We use the next-generation matrix method to compute the $\Re_0$ when the entire population is susceptible and has no contact and infection compartment. we will consider two different types of equations:
\begin{equation}
	\begin{cases}
    	\begin{split}
		\frac{dE}{dt} &=  \beta S E-\alpha \delta E-(1-\alpha)\delta E-\nu E,\\
	\frac{dI}{dt} &=\alpha \delta E-k I-\gamma \eta I-\nu I.
    	\end{split}
 \end{cases}
	\label{eq:sffvvvystem22}.
\end{equation}
From the equation above, we obtained
\begin{equation}\label{22}
F=	\begin{pmatrix}
		\beta S E\\
		0
	\end{pmatrix},
\end{equation}
and 
\begin{equation}\label{12}
V=	\begin{pmatrix}
		(\delta +\nu )E\\
		-\alpha \delta E+(k +\gamma \eta +\nu )I\\
	\end{pmatrix}.
\end{equation}

The Jacobian matrices at MFE of the matrices in equations \ref{22} and \ref{12} is given as:

\begin{equation}\label{23}
J_F=	\begin{pmatrix}
	\frac{\beta\Phi}{\nu}&0 \\
		0&0
	\end{pmatrix},
\end{equation}
and 
\begin{equation}\label{24}
J_V=	\begin{pmatrix}
		\delta+\nu&0\\
		\alpha\delta&\eta \gamma+k+v
	\end{pmatrix},
\end{equation}
thus
\begin{equation}
	J_{F}\cdot  J_{V}^{-1}=	\begin{pmatrix}
	\frac{\Phi\beta}{\nu(\delta+\nu)}&0\\
		0&0
	\end{pmatrix}.
\end{equation}
Hence, the threshold parameter $\Re_{0}$ is obtained by taking the largest eigenvalue 
\begin{equation}
	\Re_{0}=\frac{\beta\Phi}{\nu(\delta+\nu)}
\end{equation}
\begin{remark}
 If $\Re_{0}>1$, the infection is expected to spread in the population. The infection will die out if $\Re_{0}<1$.  
\end{remark}

\subsection{Local Stability Analysis of Mob Free Equilibrium}
The analysis of the model’s equilibrium stability shows the following results.
\begin{theorem}
The MFE point is locally asymptotically stable if $\Re_{0}<1$ and unstable otherwise.
\end{theorem}
\proof We examine the Jacobian of model \ref{eq:system22}, given by:
	\begin{equation}\label{ja}
	J=	\begin{pmatrix}
		-\beta E-\nu&-\beta S&0&0&\Psi\\
		\beta  E&\beta S-\alpha\delta-\delta(1-\alpha)-\nu&0&0&0\\
		0&\alpha\delta&-\eta\gamma-k-\nu&0&0\\
		0&\delta(1-\alpha)&k&-\eta(1-\gamma)-\nu&0\\
		0&0&\eta\gamma&\eta(1-\gamma)&-\Psi-\nu
	\end{pmatrix},
\end{equation}
Evaluating equation (\ref{ja}) at the MEF point (\ref{MFE}), we obtain
	\begin{equation}
		J=	\begin{pmatrix}
		-\nu&\frac{-\beta\Phi}{\nu}&0&0&\Psi\\
		0&\frac{-\beta\Phi}{\nu}-\alpha\delta-\delta(1-\alpha)-\nu&0&0&0\\
		0&\alpha\delta&-\eta\gamma-k-\nu&0&0\\
		0&\delta(1-\alpha)&k&-\eta(1-\gamma)-\nu&0\\
		0&0&\eta\gamma&\eta(1-\gamma)&-\Psi-\nu
	\end{pmatrix}.
\end{equation}
	From the above Jacobian matrix, the negative eigenvalues are
	$\lambda_{1}=-\nu,  \lambda_{2}=-\frac{-\Phi\beta+\delta\nu-\nu^{2}}{\nu},\\\lambda_{3}=-\Psi-\nu$, $\lambda_{4}=-\eta\gamma-k-\nu$  and $\lambda_{5}= \eta\gamma-\eta-\nu$.
 Thus, the mob-free equilibrium is asymptotically stable locally.	
	\subsection{Existence of Endemic Equilibrium State}
The endemic equilibrium point of the model \ref{eq11} is identified when $ S\neq E\neq I\neq A\neq R\neq 0$,
 denoted by $E^{\star} = (S^{\star}, E^{\star}, I^{\star}, A^{\star}, R^{\star}) \neq 0$ and can be obtained by setting each equation of the ODEs system to zero, i.e.,
	\begin{equation}
		\frac{dN}{dt}=	\frac{dS}{dt}=	\frac{dE}{dt}=	\frac{dI}{dt}=	\frac{dA}{dt}=	\frac{dR}{dt}=0.
	\end{equation}
	Then, we obtain
	\begin{equation}\label{EPE}
 	\begin{cases}	\begin{split}
S^{\star}&= \frac{\delta+\nu}{\beta},\\
	E^{\star}&=\frac{(\Psi+\nu)(-\eta\gamma+\eta+\nu)(\eta\gamma+k+\nu)\phi_{1}}{\beta \nu\phi{2}},\\
	I^{\star}&=-\frac{\alpha\delta(\Psi+\nu)(-\eta\gamma+\eta+\nu)\phi_{1}}{\beta \nu \phi_2}, \\
	A^{\star}&= \frac{\delta(\Psi+\nu)(\alpha\eta\gamma+\alpha \nu-\eta \gamma-k-\nu)\phi_{1}}{\beta \nu \phi_2},\\
	R^{\star}&=-\frac{\delta\eta (2\alpha\gamma \nu-\alpha \nu-\eta \gamma^{2}+\eta\gamma-\gamma k- \gamma \nu+k+\nu)\phi_{1}}{\beta \nu \phi_2}, 
    	\end{split}
 	\end{cases}
\end{equation}	
where 
\begin{eqnarray*}
    \phi_{1}&=&-\Phi\beta+\delta \nu+\nu^{2},\\
    \phi_{2}&=&-2\Phi \alpha\delta\eta \gamma+\Psi\alpha\delta\eta+\Psi\delta\eta\gamma+\Psi\delta k+\Psi \delta v-\Psi \eta^{2}\gamma^{2}+\Psi\eta^{2}\gamma-\Psi\eta\gamma k+\Psi\eta k\notag\\&+&\Psi \eta \gamma+\Psi\eta \nu+\Psi \nu^{2}-\delta\eta^{2}\gamma^{2}+\delta \eta^{2}\gamma-\delta\eta\gamma k+\delta \eta k+\delta \eta \nu+\delta k \nu+\delta \nu^{2}-\eta^{2}\gamma^{2}\nu\notag\\&+&\eta^{2}\gamma \nu-\eta \gamma k \nu+\eta k \nu+\eta \nu^{2}+k \nu^{2}+\nu^{3}.
\end{eqnarray*}

The system equation (\ref{EPE}) demonstrates that if $\Re_{0}>1$, then the endemic equilibrium  $ E^{\star} = (S^{\star}, E^{\star}, I^{\star}, A^{\star}, R^{\star})\in \Sigma$. In this analysis, we examine the equilibrium states of a SEIAR model by varying the transmission parameter $\beta$. The model utilizes  parameter values: $\Phi = 0.5, \beta = 0.3, \alpha = 0.5, \delta = 0.2, k = 0.1, \gamma = 0.7, \nu = 0.1, \Psi = 0.3$, and $\eta = 0.5$. These parameters define the rates of transitions between susceptible, exposed, against (non-spreading), infected, and recovered individuals. For each value of $\beta$, the equilibrium values of these states $S^{\star}, E^{\star}, A^{\star}, I^{\star}, R^{\star}$ were computed. Figure \ref{SEIjhhj,jjmmREq} shows how changes in $\beta$ affect the equilibrium distribution of the population.
\begin{figure}[!ht]
	\centering
	\includegraphics[width=.5\textwidth]{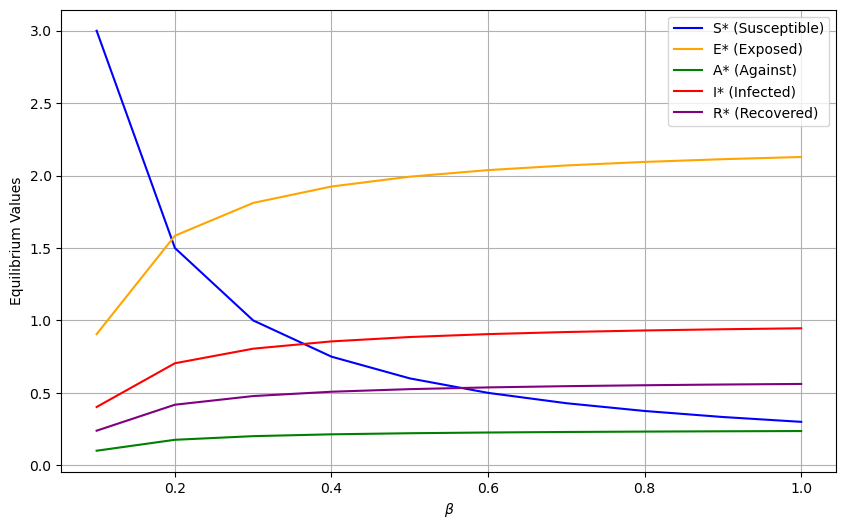} % Adjust the height as needed
	\caption{Global Stability Analysis of Endemic Equilibrium in the SEAIR Model of the COVID19 model.}
	\label{SEIjhhj,jjmmREq}
\end{figure}
  %%%%%%%%%%%%%%%%%%%%%%%%%%%%%%%%%%%%%%%%%%%%%%%
%%%%%%%%%%%%%%%%%%%%%%%%%%%%%%%%
\subsection{Sensitivity analysis}
In this section, we performed a sensitivity analysis to illustrate the effect of each parameter on the mob transmission. The sensitivity indices concerning the parameter values are given in the form of
\begin{eqnarray}
	\Pi_{\rho}^{\Re_{0}}=\frac{d\Re_{0}}{d\rho}\times\frac{\rho}{\Re_{0}},
\end{eqnarray}
where $\rho$ denotes all the basic parameters and 
\begin{equation}\label{eqf}
	\Re_{0}=\frac{\beta\Phi}{\nu(\delta+\nu)}
\end{equation}
\begin{eqnarray}
	\Pi_{\beta}^{\Re_{0}}=\frac{d\Re_{0}}{d\beta}\times\frac{\beta}{\Re_{0}}=1>0,
\end{eqnarray}
\begin{eqnarray}
	\Pi_{\Phi}^{\Re_{0}}=\frac{d\Re_{0}}{d\Phi}\times\frac{\Phi}{\Re_{0}}=1>0.
\end{eqnarray}
\begin{eqnarray}
	\Pi_{\nu}^{\Re_{0}}=\frac{d\Re_{0}}{\nu}\times\frac{\nu}{\Re_{0}}=\frac{-\delta-2\nu}{\delta+\nu}<0.
\end{eqnarray}
\begin{eqnarray}\label{eql}
	\Pi_{\delta}^{\Re_{0}}=\frac{d\Re_{0}}{d\delta}\times\frac{\delta}{\Re_{0}}=-\frac{\delta}{\delta + \nu}<0.
\end{eqnarray}
The numerical values indicating the relative importance of $\Re_0$ parameters are shown in equations (\ref{eqf})-(\ref{eql}). As one can see, $\Phi$ and $ \beta$ affect the mob stability. Therefore, by increasing $\Phi$ and $ \beta$ by 1\%, $\Re_0$ would increase by 1\% in influence. Some parameters have a positive relation, while others have a negative relation.
%%%%%%%%%%%%%%%%%%%%%%%%%%%%%%%%%%%%%%%%%%%%%%
\begin{figure}[!ht]
	\centering
	\includegraphics[width=0.6\textwidth]{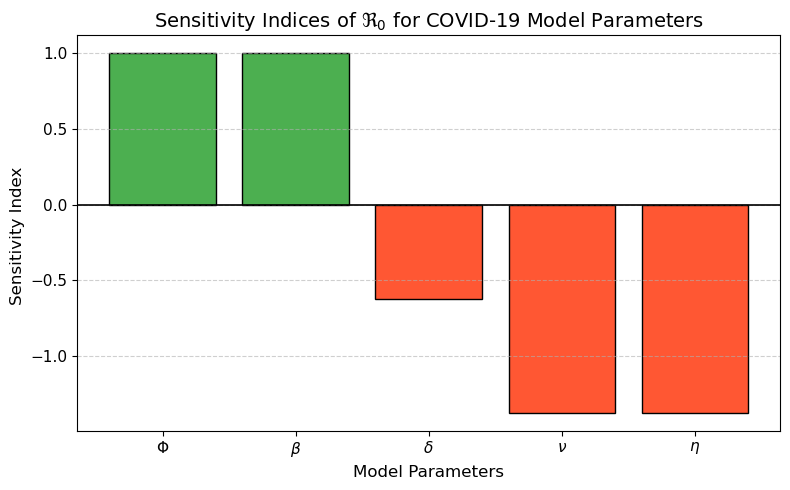} % Adjust the height as needed
	\caption{The figure shows the sensitivity indices of $\Re_0$ for the dependent parameters of the COVID19 model.}
	\label{SEIjjjmmREq}
\end{figure}
  %%%%%%%%%%%%%%%%%%%%%%%%%%%%%%%%%%%%%%%%%%%%%%%
A negative relation suggests that increasing the metrics values would help reduce mob event brutality. At the same time, a positive relationship indicates that an increase would greatly influence the frequency of the mob event in the values of those parameters. Figure \ref{SEIjjjmmREq} shows the sensitivity indices for the dependent parameters of the basic reproduction number.
\section{Numerical simulations and discussions}\label{sec3}
	This section proposes some numerical tests to validate the theoretical results. We apply the SEIAR model to find the numerical solution in the computation. The numerical experiment utilizes a Twitter dataset related to the COVID-19 epidemic from April 2020 to June 2020. The dataset consisted of approximately 2,35,240 tweets interacting with the COVID-19 epidemic during the epidemic life cycle.  
    
    \subsection{Fitting Data}
	This section aimed to study individual changes in various compartments of the SEIAR  model for the Twitter dataset. A least squared measure \cite{wewer} was used as given by:
	{\begin{equation*}
		\Theta = \text{argmin}_{\varphi}\left(\sum_{i=0}^{\text{n}}\left(I_{cu}(x_{i})-\underbrace{I_{vcu}(x_{i})}_\text{tweet}\right)^{2}\right),
	\end{equation*}
	where $\Theta$ represents the vector containing evaluated parameter, $\varphi$ symbolizes the parameter space, $x_{\text{n}}$ is the latest date considered in the analysis, $x_{i}$ represents the date, $I_{cu}(x_{i})$ signifies the cumulative function incidence up to $x_{i}$, and $I_{vcu}(x_i)$ represents the cumulative function incidence according to the model up to $x_{i}$. In order to evaluate the SEIAR models, Figures \ref{NEWSjjEAIR} show how sufficiently the SEIAR models fit the COVID-19 on the Twitter data, along with the 2-norm relative error
		$E_{- \text{rel}} = \frac{\left\|I_{cu}(x_{i}) - I_{vcu}(x_{i})\right\|_{2}}{\left\|I_{vcu}(x_{i})\right\|_{2}},$
	and the mean absolute error defined by $ \frac{\left\|I_{cu}(x_{i}) - I_{vcu}(x_{i})\right\|_{1}}{n}.$	
	The numerical experiments indicate that the SEIAR model accurately fits Twitter data for the COVID-19 virus. The lower error in the SEIAR model indicates a precise representation of Twitter data. The SEIAR model invariably captures the initial spread more influential on Twitter, a phenomenon attributed to a delay as individuals in the `Exposed' class take time before transmitting their posts.
%%%%%%%%%%%%%%%%%%%%%%
\begin{figure}[!ht]
    \centering
    % First row of images
    \begin{subfigure}{0.45\textwidth}
        \centering
        \includegraphics[width=\textwidth]{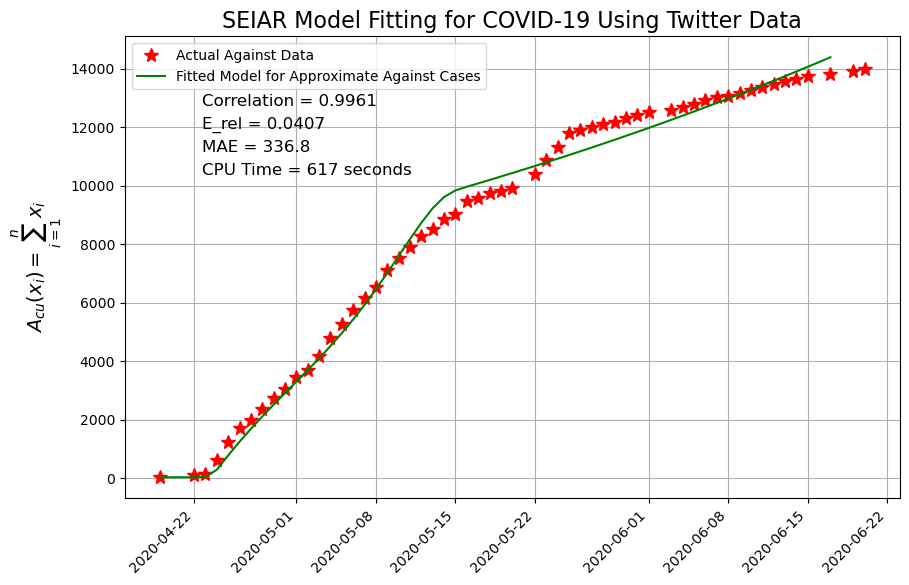}
        \caption{Against (Non-spreading) (A)}
        \label{fig:surface}
    \end{subfigure}
    \hfill
    \begin{subfigure}{0.45\textwidth}
        \centering
        \includegraphics[width=\textwidth]{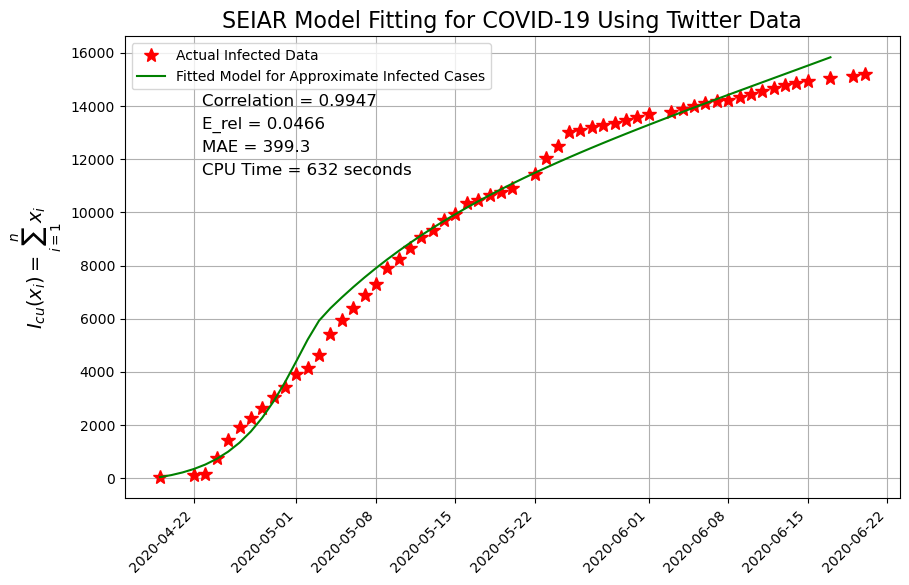}
       \caption{Infected (I)}
        \label{fig:contour1}
    \end{subfigure}
   \caption{Best fit modeling for mob Dynamics in social media networks.  Figure (a) shows the estimated against case data fitting with the real against case data of the SEAIR model, and figure (b) shows the estimated infected case data fitting with the real infected case data of the SEAIR model}
    \label{NEWSjjEAIR}
\end{figure}

%%%%%%%%%%%%%%%%%%%%%%
%Basic Reproductive Number (R0)
	\subsection{Analyzing the Reproduction Number $\Re_0$}
		The basic reproduction number is the mean of secondary infections one infected individual generates within a fully susceptible population. The application lies in considering whether a new infectious disease has the prospect of growing among a population. This exam utilizes suitable contour plots to present the effect of essential parameters in SEIAR model \ref{eq:system22} on the reproductive number $\Re_0$.  Fig. \ref{SEIjjjREq} displays how $\Re_0$ changes with the effective contact rates $\beta$ and $\Phi$ in the SEIAR model. We find that $\Re_0$ has a lower value when  $\beta$ and $\Phi$ drop below $1$.
%%%%%%%%%%%%%%%%%%%%%%
\begin{figure}[!ht]
    \centering
    % First row of images
    \begin{subfigure}{0.45\textwidth}
        \centering
        \includegraphics[width=\textwidth]{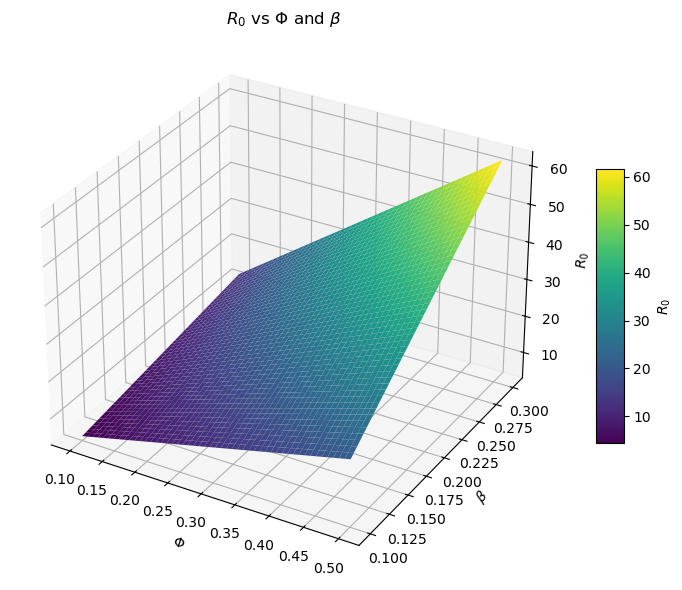}
        %\caption{}
        \label{fig:surface}
    \end{subfigure}
    \hfill
    \begin{subfigure}{0.45\textwidth}
        \centering
        \includegraphics[width=\textwidth]{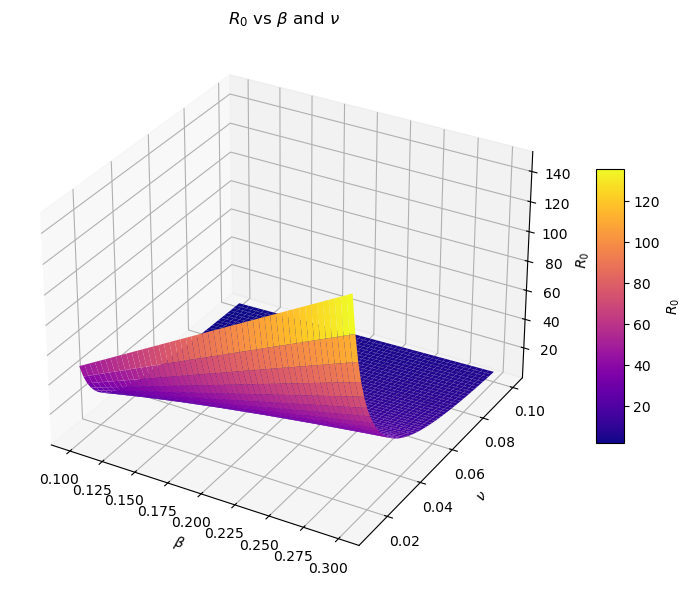}
        %\caption{}
        \label{fig:contour1}
    \end{subfigure}
    % Second row of images
    \vspace{0.5cm}
    \begin{subfigure}{0.45\textwidth}
        \centering
        \includegraphics[width=\textwidth]{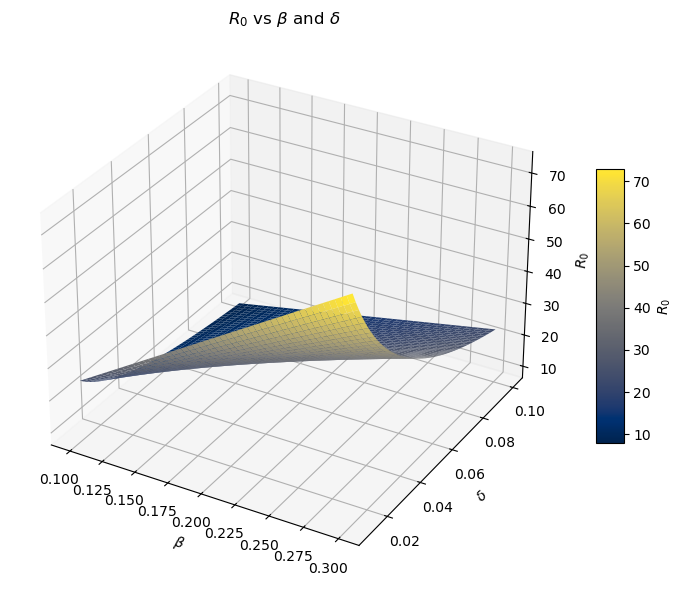}
        %\caption{}
        \label{fig:contour2}
    \end{subfigure}
    \hfill
    \begin{subfigure}{0.45\textwidth}
        \centering
        \includegraphics[width=\textwidth]{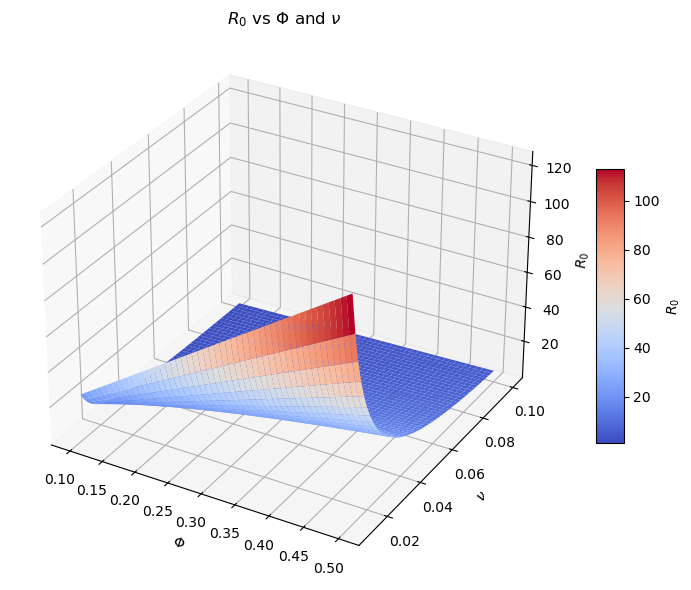}
        %\caption{}
        \label{fig:contour3}
    \end{subfigure}
    \caption{Dynamics of $\Re_0$ on different parameters.}
    \label{SEIjjjREq}
\end{figure}

%%%%%%%%%%%%%%%%%%%%%%
%%%%%%%%%%%%%%%%%%%%%%
\subsection{Caputo fractional derivative $^{C}_{0}D^{\alpha}_{t}\varPsi(t)$}
		Many natural processes, such as the spread of diseases, show time-memory effects, which provide insights into biological systems with long-term influences. Fractional derivative models help address these effects because they include time-dependent factors. In this study, we use the Caputo fractional derivative, which has the same initial conditions as standard derivatives, making it easier to work with. Using this approach, we adjust the SEIAR model (\ref{FRAC}) by incorporating the Caputo fractional time derivative.
        \begin{definition}[See \cite{podlubny1998fractional,baleanu2010new}]
		\label{Definition2.1}
		Let {{$\varPsi:\Re^{+}\longrightarrow \Re$}}, the Caputo derivative {{$^{C}_{0}D^{\alpha}_{t}\varPsi(t)$}} is define as 
	{{\begin{equation}\label{Caputo}
			^{C}_{0}D^{\alpha}_{t}\varPsi(t)=\frac{1}{\Gamma(k-\alpha)}\int^{t}_{0}(t-s)
			^{k-\alpha-1}\varPsi^{(k)}(\phi){\rm d}\phi,
		\end{equation}}}
  where $\quad t>0, k-1<\alpha\leq  k, k \in \mathbb{N}.$
	\end{definition}
	\begin{lemma}[See \cite{podlubny1998fractional}]
		\label{Lemma2.3}
		Assume that  {{$\varPsi\in C[0,\delta]$}}.
		The approximation of the fractional differential equation
		{{\begin{equation*}
			\left\{
			\begin{array}{llllllllll}
				_{0}^{C}D^{\alpha}_{t}\varPsi(t)=\Upsilon(t,\Psi(\phi)),
				\quad t\in[0,\delta], \cr
				\varPsi(0) = \varPsi_{0},
			\end{array}
			\right.
		\end{equation*}}}
		is given by
		{{$$
		\varPsi\left( t\right) -\varPsi\left( 0\right) = \frac{1}{\Gamma({\alpha})}
		\int ^t_0 \Upsilon (\phi,\varPsi(\phi))(t-\phi)^{\alpha-1} d\phi.
		$$}}
	\end{lemma}

        The following system of ODEs represents the SEIAR model :
\begin{equation}
			\begin{cases}
            	\begin{split}
				^{C}_{0}D^{\alpha}_{t}  S &=\Phi+\Psi R-\beta S  E -\nu S\\
				^{C}_{0}D^{\alpha}_{t}  E &= \beta S E-\alpha \delta E-(1-\alpha)\delta E-\nu E \\
				^{C}_{0}D^{\alpha}_{t} I &=\alpha \delta E-k I-\gamma \eta I-\nu I\\
				^{C}_{0}D^{\alpha}_{t} A &= (1-\alpha)\delta E+k I-(1-\gamma)\eta A-\nu A\\
				^{C}_{0}D^{\alpha}_{t} R &=(1-\gamma)\eta A+\gamma \eta I-\Psi R-\nu R.  
                \end{split}
			\end{cases}
			\label{FRAC}
\end{equation}

We simulate the spread of COVID-19 using the SEIAR models (\ref{FRAC}) with Caputo fractional derivatives. The starting conditions are $S(0)=400, E(0)=10,I(0)=0, A(0)=0 \mbox{ and } R(0)=0$. The model parameters are based on Table \ref{tab:seiz-parameters}, and the simulations are performed in Python using the Caputo fractional method as shown in Figure \ref{NEWllSjjEAIR}. In Figure \ref{NEWllSjjEAIR}, Twitter data for Infected (I) and Against (A) indicate different trends for various $\alpha$ values. Infected tweets peak and then stabilize, reflecting ongoing activity before stopping. Against cases, increase, implying changes in the spread of COVID-19 content as individuals reduce or stop posting and may repost existing content. The choice of the fractional order $\phi$ significantly affects the dynamics of the SEIAR model. Lower values introduce long-memory effects, leading to delayed peaks and prolonged infections, whereas higher values result in quicker spread and lower infection peaks. These insights can help policymakers design better strategies for disease control based on the underlying fractional dynamics of the epidemic.
%%%%%%%%%%%%%%%%%%%%%%
\begin{figure}[H]
    \centering
    \begin{minipage}{0.45\textwidth}
        \centering
        \includegraphics[width=\textwidth]{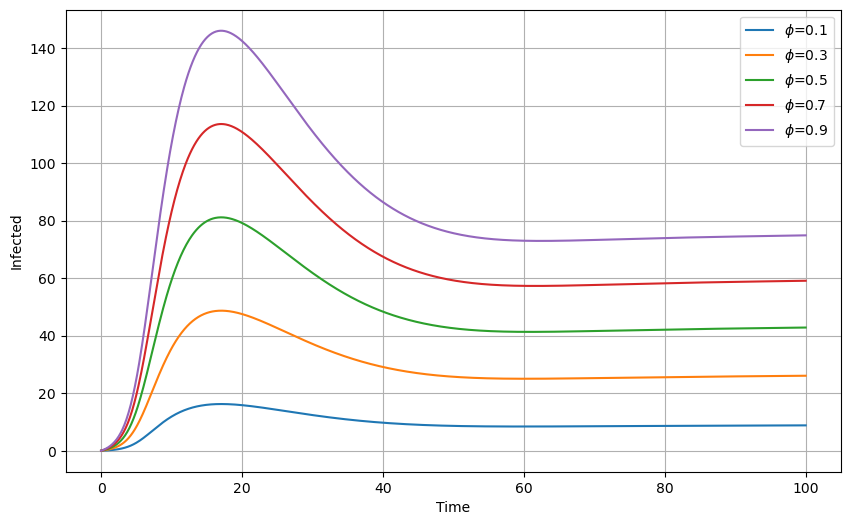}
    \end{minipage}%
    \hfill
    \begin{minipage}{0.45\textwidth}
        \centering
        \includegraphics[width=\textwidth]{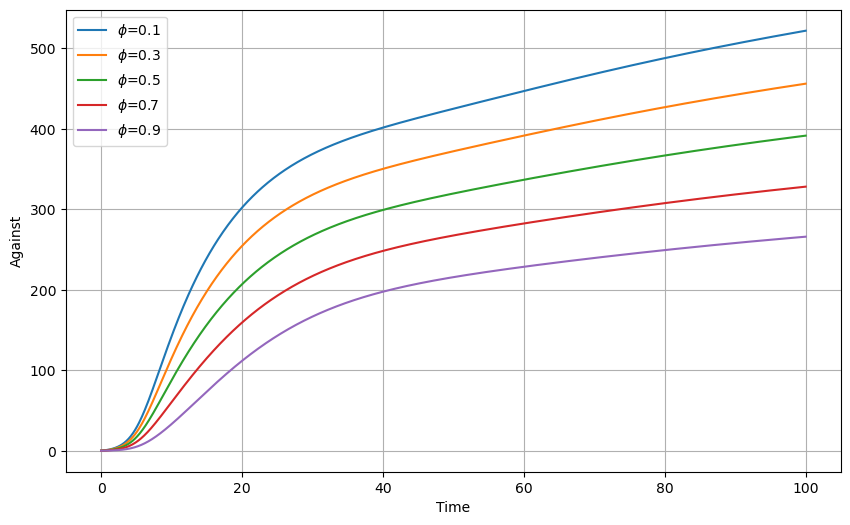}
    \end{minipage}
    \caption{computational simulation of the SEIAR model using the Caputo
    differential operator with different fractional orders $\phi$.}
    \label{NEWllSjjEAIR}
\end{figure}
%%%%%%%%%%%%%%%%%%%%%%
%%%%%%%%%%%%%%%%%%%%%%
\subsection{Simulation of The SEIAR Model}
We discuss the numerical simulations to examine the infected by the proposed model. We consider simulation results from different iterations and dynamical behavior in the SEIAR model \ref{mmm}. As results depicted in \ref{fig:simulation_dynamics} confirm, the numbers of infected individuals grew over time when exposed, and the numbers slowed down. Intuitively, contact tracing in the network and isolate the infectious nodes. Then, it will reduce the rate of outbreaks as much as possible. 

Figure \ref{fig:simulation_dynamics} shows the effect of mobility simulation in the population and what is the change in the behavior of spread outbreaks with parameters value ($\Phi =2, \beta =0.14, \epsilon= 0.26, \Phi =0.0074, \delta= 0.10, \mu =0.10, v = 0.05)$. 
In general, mobility can be defined as a network of
interacting communities where the connections and the interrelated intensity represent the movement among them \cite{balcan2009multiscale}.
%Model
The model is constructed with the Mobility equation and Monte Carlo simulation  as follows:\begin{equation}
			\begin{cases}
				\frac{dS}{dt} &=\Phi+\Psi R-\beta S  E -\nu S+\sum_{j}T_{ji}S_{j}-\sum_{j}T_{ij}S_{i},\\
				\frac{dE}{dt} &= \beta S E-\alpha \delta E-(1-\alpha)\delta E-\nu E+\sum_{j}T_{ji}E_{j}-\sum_{j}T_{ij}E_{i}, \\
				\frac{dI}{dt} &=\alpha \delta E-k I-\gamma \eta I-\nu I+\sum_{j}T_{ji}I_{j}-\sum_{j}T_{ij}I_{i},\\
				\frac{dA}{dt} &= (1-\alpha)\delta E+k I-(1-\gamma)\eta A-\nu A+\sum_{j}T_{ji}A_{j}-\sum_{j}T_{ij}A_{i},\\
				\frac{dR}{dt} &=(1-\gamma)\eta A+\gamma \eta I-\Psi R-\nu R+\sum_{j}T_{ji}R_{j}-\sum_{j}T_{ij}R_{i},  
            \end{cases}
			\label{mmm}
\end{equation}
where $\epsilon(t)$ is the time-varying contact rate that mob dynamics can adjust via Monte Carlo simulations, the mobility term $\sum_{j} T_{ji} S_{j}$ denotes people joining region $i$. In contrast, $\sum_{j} T_{ij} S_{j}$ denotes people exiting region $i$ and $\epsilon(t)=\epsilon(0)\cot (1-M_{t})-M_{t} $ is random factor between $0$ and $1.5.$ 
%%%%%%%%%%%%%%%%%%%%%%%%%%%%%%%%%%
%%%%%%%%%%%%%%%%%%%%%%
\begin{figure}[!ht]
    \centering
    % First row of images
    \begin{subfigure}{\textwidth}
        \centering
        \includegraphics[width=.8\textwidth]{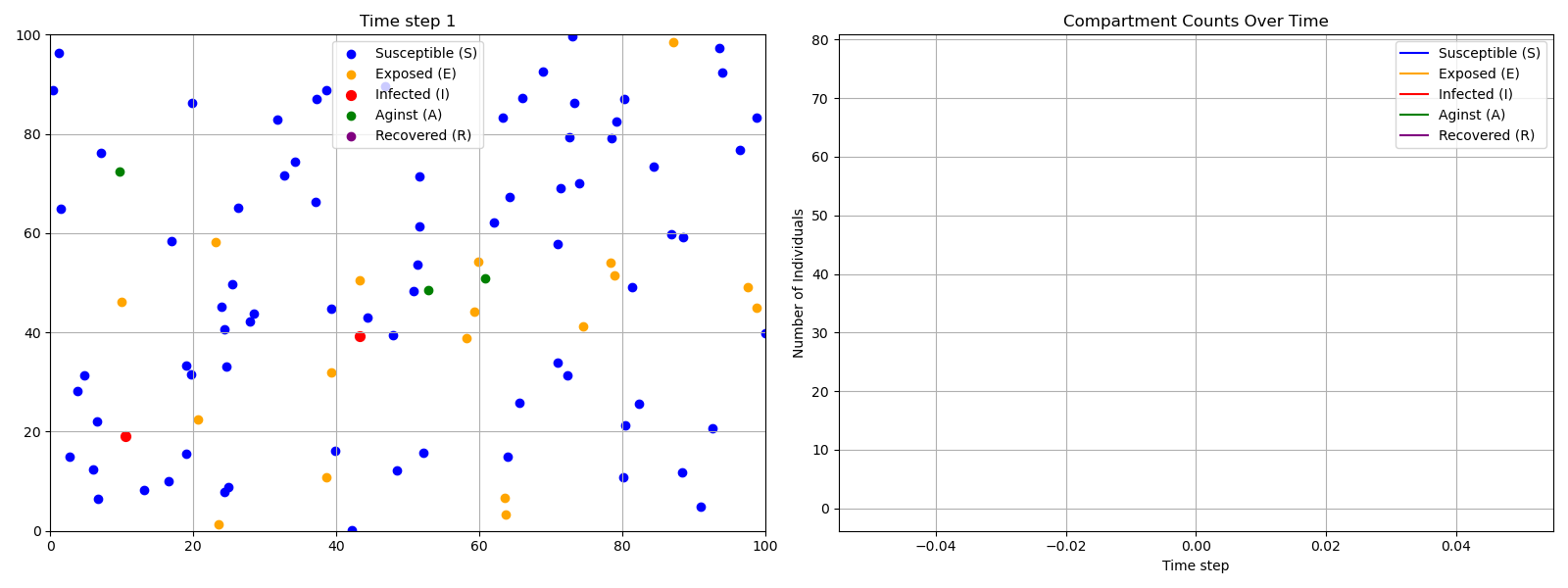}
        \label{fig:surface}
    \end{subfigure}
    \hfill
    \begin{subfigure}{\textwidth}
        \centering
        \includegraphics[width=.8\textwidth]{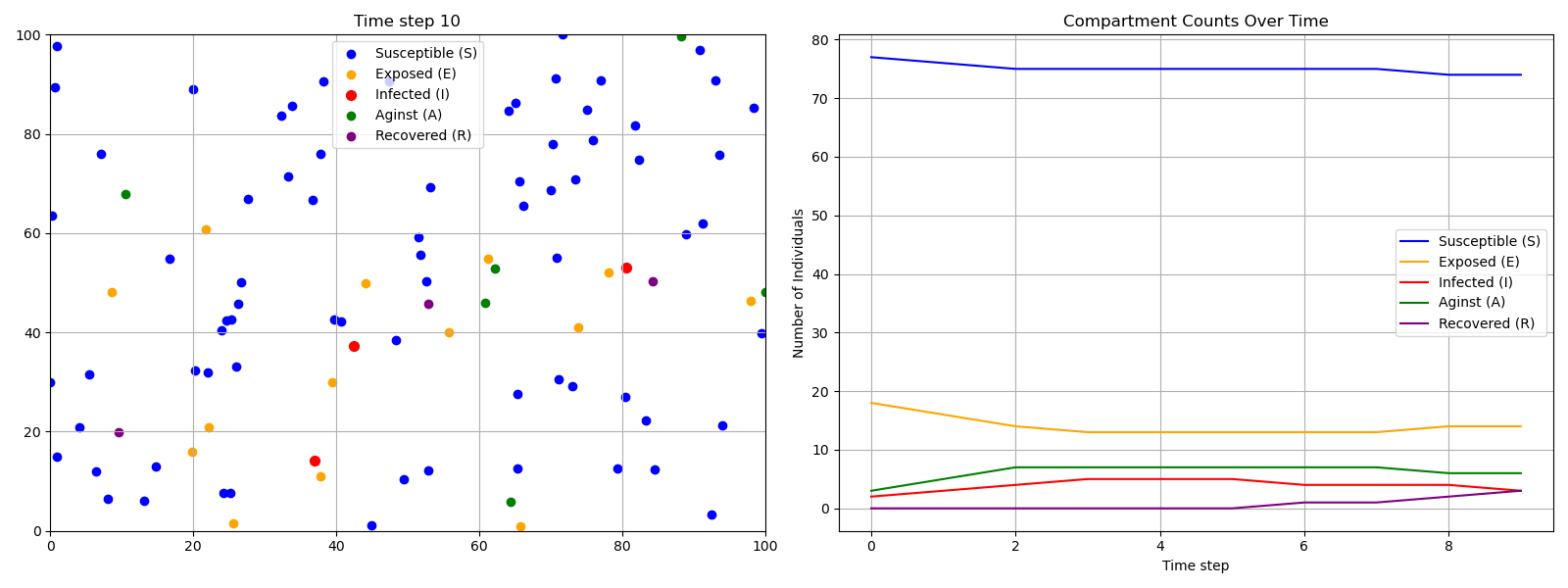}
        \label{fig:contour1}
    \end{subfigure}
    \hfill
    \begin{subfigure}{\textwidth}
        \centering
        \includegraphics[width=.8\textwidth]{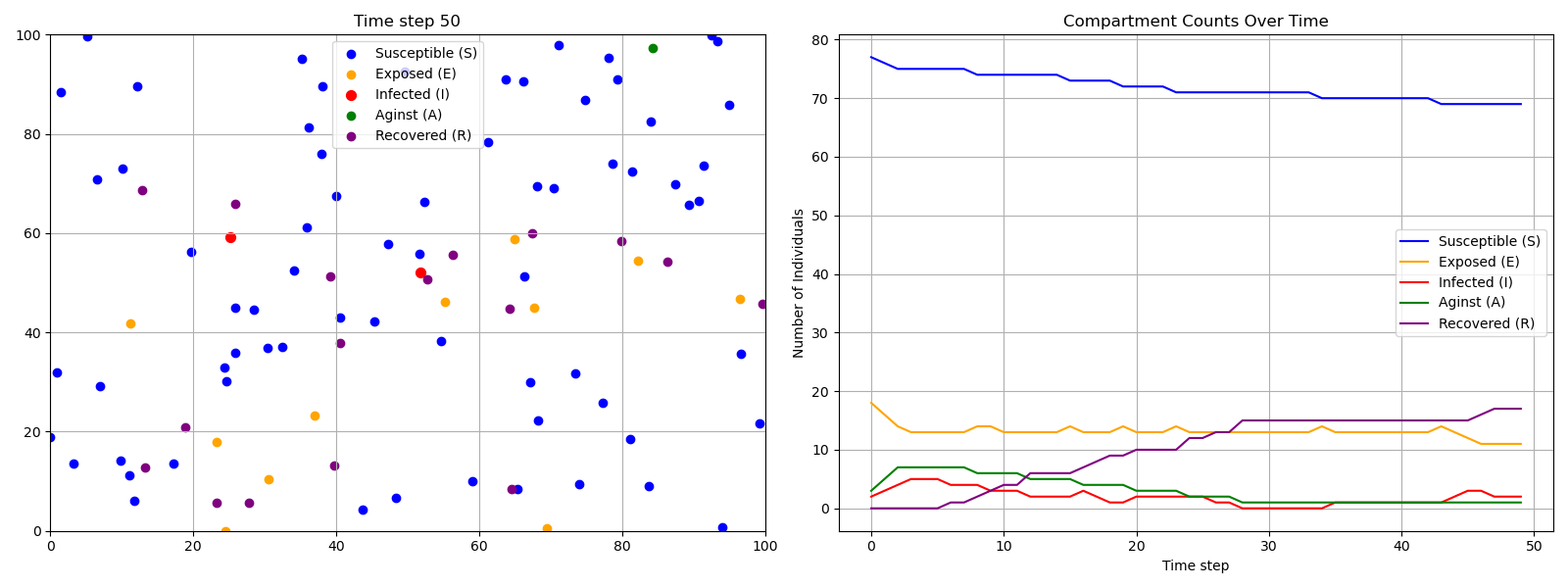}
        \label{fig:contour2}
    \end{subfigure}
    \hfill
    \begin{subfigure}{\textwidth}
        \centering
        \includegraphics[width=.8\textwidth]{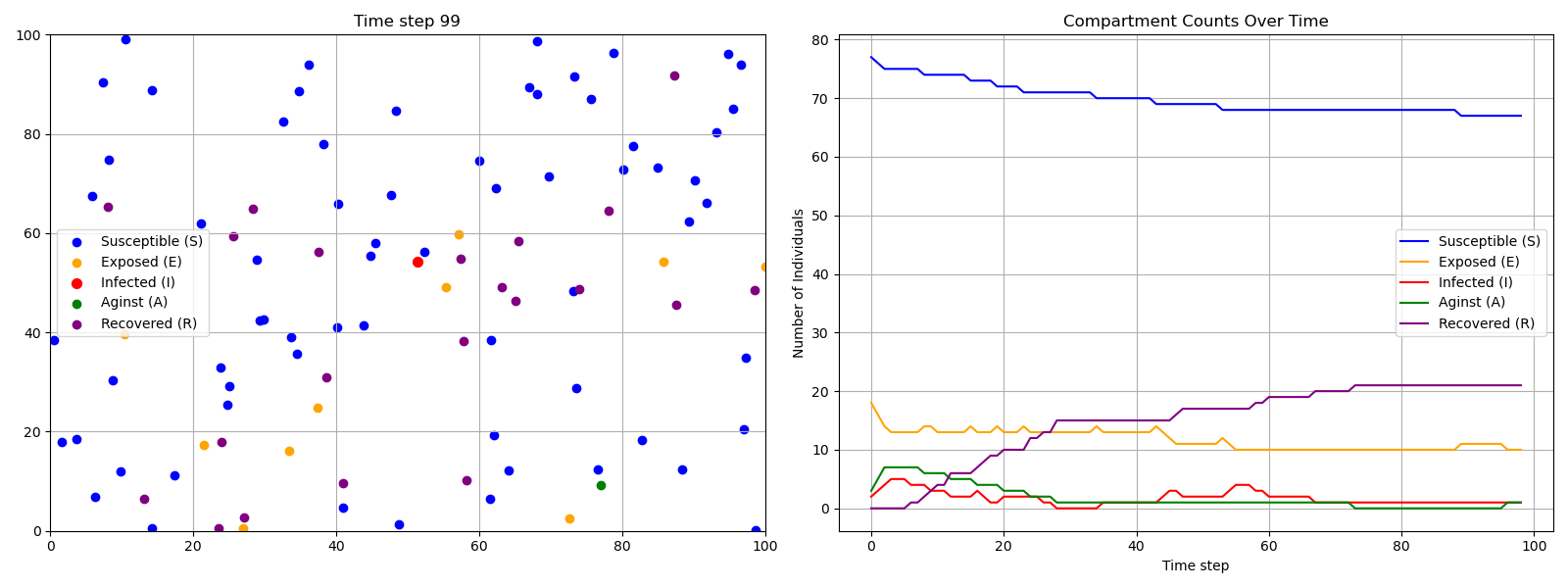}
        \label{fig:other}
    \end{subfigure}
    \caption{Simulation of the proposed model \ref{mmm} showing mobility and disease dynamics over time.}
    \label{fig:simulation_dynamics}
\end{figure}

%%%%%%%%%%%%%%%%%%%%%%%%%%%%%%%%%%%%%%%%%%%%%%%%%%%%%%%%%%
\section{Conclusions}\label{sec4}
In this paper, we developed a mathematical model to illustrate the dynamics of social media mob events. Our analysis reveals that the disease-free equilibrium of the model is locally asymptotically stable when the associated $\Re_0$ is less than one, and it becomes unstable otherwise. We computed the basic reproduction number $\Re_0$ and examined the stability of the equilibrium points in detail. Based on our findings, we suggest that stakeholders and policymakers adopt an integrated strategy to mitigate the effect of social media events on the population.

Furthermore, our study explores COVID 19 spread on Twitter using a Caputo derivative, revealing insights into dynamics and stability. As transmission and recovery rates decrease below 1 in the SEIAR model, the reproductive number ($\Re_0$) indicates potential control over virus spread.  The main goal is to aid policymakers in combating misinformation by targeting interventions toward skeptical users and understanding mob events' impacts on public opinion and decision-making processes
\section*{Compliance with Ethical Standards}
Conflict of interest: No conflict of interest exists in the submission of this manuscript.
%%%%%%%%%%%%%%%%%%%%%
\bibliographystyle{plain}
	\bibliography{REF}		
%%%%%%%%%%%%%%%%%%%%%%
\end{document}